\documentclass[preprint%
 reprint,
 amsmath,amssymb,
 aps,
]{revtex4-2}

\usepackage{graphicx}
\usepackage{dcolumn}
\usepackage{bm}
\usepackage{float} 
\usepackage{xcolor}

\usepackage{lipsum} 
\usepackage{subfiles}

\newcommand*{\affmark}[1][*]{\textsuperscript{#1}}

\begin{document}


\title{Measurement of the Isomeric Yield Ratio for Zirconium-88\\ Thermal Neutron Absorption} 

\author{Isaac~Kelly\affmark[1]}
\author{Will~Flanagan\affmark[1,2]}\email{wflanagan@udallas.edu}
\author{Jacob~Moldenhauer\affmark[1]}
\author{William~Charlton\affmark[2]}
\author{Joseph~Lapka\affmark[2]}
\author{Donald~Nolting\affmark[2]}
\affiliation{\affmark[1]Department of Physics, University of Dallas, Irving, TX, 75062, USA}
\affiliation{\affmark[2]Nuclear Engineering Teaching Laboratory, University of Texas, Austin, TX, 78758, USA}

\date{\today}

\begin{abstract}

In light of the recently observed 800,000 barn thermal neutron absorption cross section of zirconium-88, this work investigates the rate (isomeric yield ratio) of metastable zirconium-89 production and resulting implications for ongoing measurements around zirconium-88 neutron absorption. The metastable state of zirconium-89 resides at 588 keV above the ground state with a half life of 4.2 minutes. A 5 $\mu$Ci zirconium-88 sample was irradiated for 10 minutes in the core of a TRIGA Mark II nuclear research reactor and measured with a high purity germanium detector starting 3 minutes after irradiation. The isomeric yield ratio was measured to be 74.9$\pm$0.6\%.

\end{abstract}

\pacs{Valid PACS appear here}
\maketitle

\section{Introduction}

The thermal neutron absorption cross section of zirconium-88 ($^{88}$Zr) was measured to be 861,000$\pm$69,000 barns in 2019 when a 10 barn cross section was expected \cite{Shusterman2019}. This value was measured again as 804,000$\pm$63,000 barns in a 2021 measurement by the same group \cite{Shusterman2021} and remains an intriguing result to the nuclear physics community. An energy-resolved transmission-based measurement has reported a 771,000$\pm$31,000 barn thermal cross section with a resonance at 0.171 eV \cite{Stamatopoulos2023}. An energy-resolved direct capture-based measurement is being performed by the authors of this study and additional collaborators at the CERN n$\_$TOF facility \cite{Weiss2015}.

\begin{figure}
    \centering
    \includegraphics[width=0.45\linewidth]{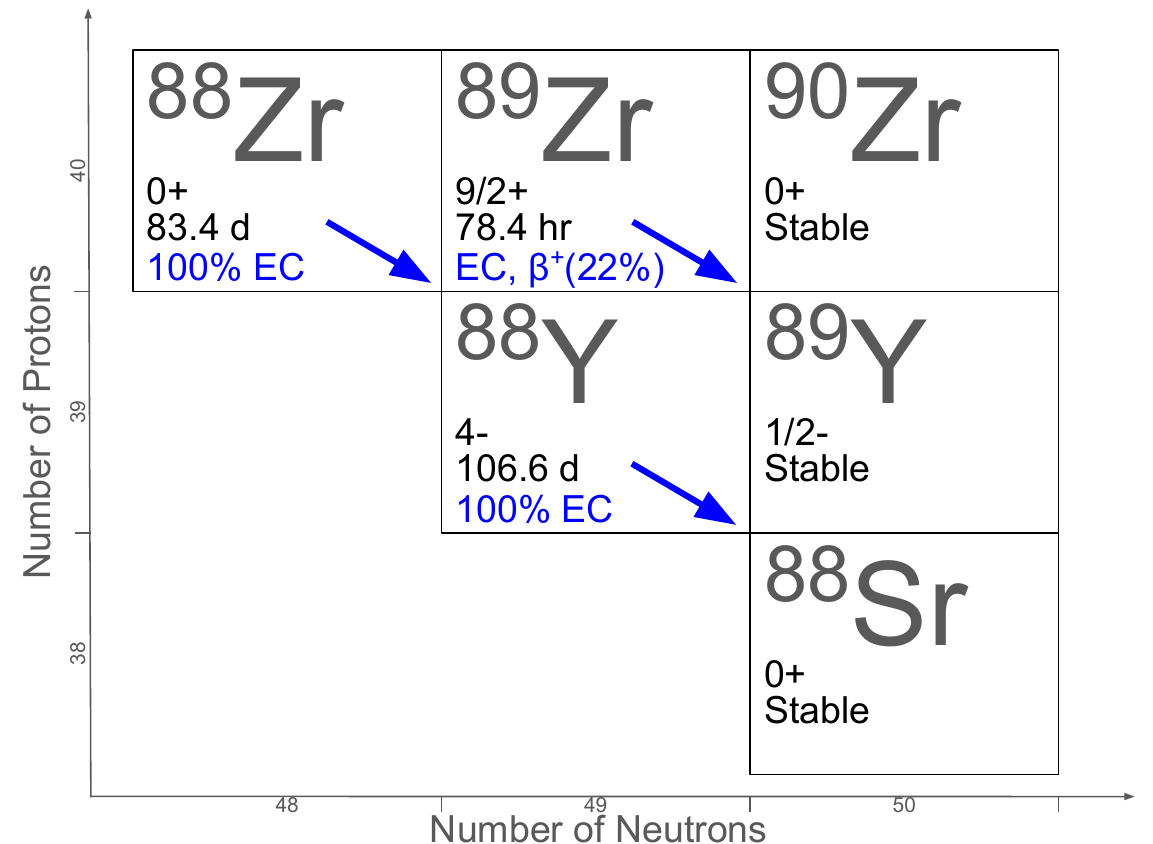} 
    \includegraphics[width=0.45\linewidth]{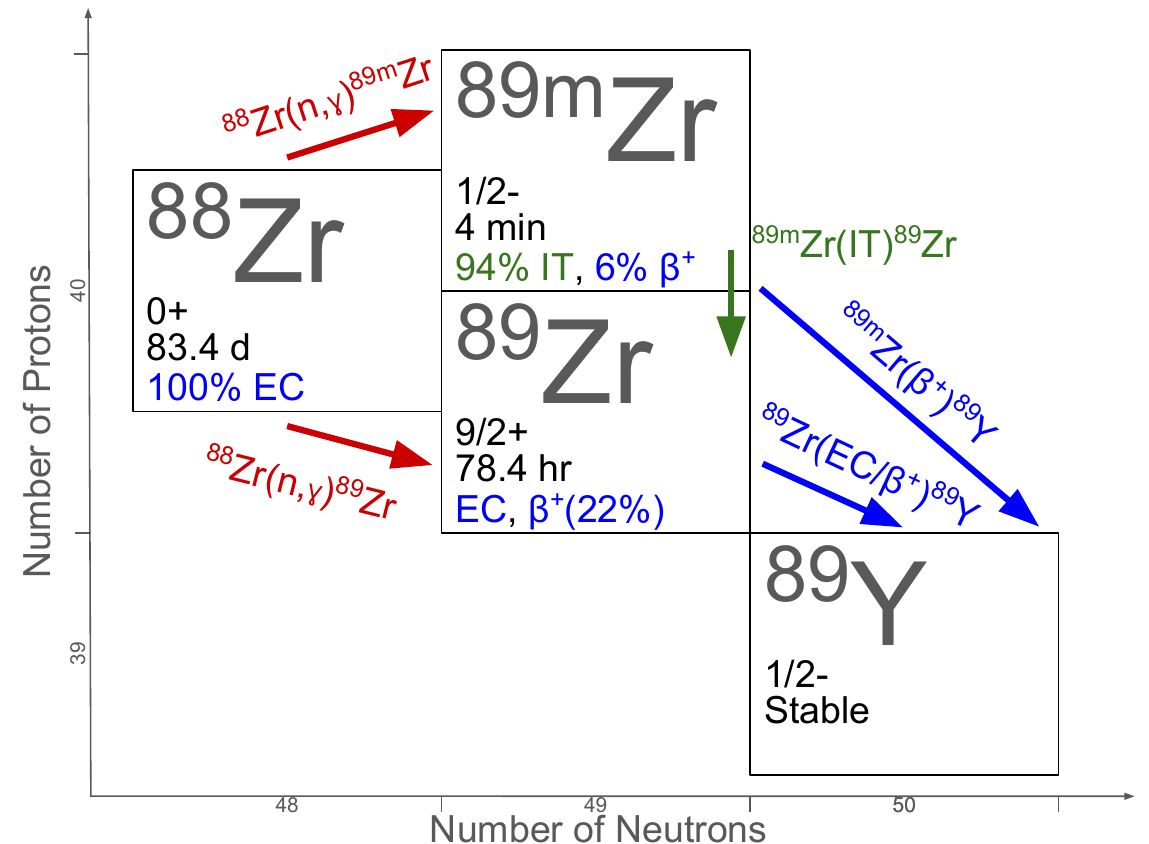}
    \caption{Left: the chart of nuclides in the vicinity of $^{88}$Zr and $^{89}$Zr. Right: the capture and decay paths of relevance to this measurement. Neutron capture reactions are shown in red. The isomeric transition of $^{89\textrm{m}}Zr$ to $^{89\textrm{g}}Zr$ is shown in green. Electron capture and positron emission are shown in blue.}
    \label{fig:nuclides}
\end{figure}

A traditional neutron capture measurement uses gamma-ray spectroscopy to measure the absolute quantities of reaction products and reactants before and after irradiation. The capture products and related decay chains must be carefully defined to maximize precision. $^{88}$Zr in neutron flux  undergoes the reaction $^{88}$Zr(n,$\gamma$)$^{89}$Zr at a high rate due to the large cross section. The product $^{89}$Zr compound nucleus has significant excess energy (9.3 MeV) which is dissipated in a cascade through nuclear energy levels to the ground state, $^{89\textrm{g}}$Zr. Some fraction of these nuclei are left in a metastable state, $^{89\textrm{m}}$Zr, at a rate defined as the isomeric yield ratio: $\textrm{IYR}\equiv\frac{^{88}Zr(n,\gamma)^{89\textrm{m}}Zr}{^{88}Zr(n,\gamma)^{89(m,g)}Zr}$.
$^{89\textrm{m}}$Zr emits a characteristic 588 keV gamma ray upon decay to ground state; its short half life (4.2 minutes) motivates counting promptly after exposure. The decay of the ground state $^{89\textrm{g}}$Zr in the sample is measured simultaneously by counting a 909 keV gamma. The relevant absorption and decay paths are summarized in Figure \ref{fig:nuclides}.

While the experimental and theory communities evaluate and explain the large $^{88}$Zr neutron absorption cross section, a measurement of the isomeric yield ratio is of interest for a few reasons. First, the IYR of the $^{88}$Zr(n,$\gamma$) reaction impacts previous measurements of the cross section when the amount of $^{89}$Zr is measured after a known fluence. Rather than an isomeric transition to the $^{89}$Zr ground state, 6.2\% of $^{89\textrm{m}}$Zr decays to $^{89}$Y \cite{VanPatter1964}. Therefore the amount of $^{89}$Zr formed by $^{88}$Zr(n,$\gamma$) depends on the fraction of reactions with $^{89\textrm{m}}$Zr as an intermediate state. Second, the IYR directly impacts the amount of prompt gamma energy released during such capture reactions (8.7 vs 9.3 MeV). Finally, the IYR of $^{89\textrm{m}}$Zr during $^{88}$Zr neutron capture is not previously measured and such a measurement allows tuning of gamma cascade generators for $^{88}$Zr(n,$\gamma$) direct capture measurements \cite{Mendoza2020}.

\section{Background}

The isomeric yield ratio of $^{89}$Zr after other reactions has been observed previously.
Mangal and Gill measured the IYR of $^{90}$Zr(n,2n)$^{89}$Zr with 14 MeV incident neutrons to be 0.72$\pm$0.08 \cite{Mangal1963}. Katz, Baker, and Montalbetti measured the IYR of $^{90}$Zr($\gamma$,n)$^{89}$Zr to have an energy dependence but remain roughly constant at 0.56$\pm$0.03 above an incident gamma energy of 16 MeV \cite{Katz1953}. Satheesh, Musthafa, Singh, and Prasad found the IYR of $^{89}$Y(p,n)$^{89}$Zr to vary between 0.2 and 0.4 at an energy range between 6 MeV and 16 MeV. \cite{Satheesh2011}. These disparate values are expected as the various reactions correspond to a range of compound nuclear energies and initial nuclear spin configurations, and none are directly applicable to the $^{88}$Zr(n,$\gamma$)$^{89}$Zr reaction which we have explored. 
Though no previous such measurements are available, the nuclear reaction model code TALYS \cite{TALYS} with $^{89}$Zr energy levels given by the RIPL-3 database \cite{RIPL3} give an IYR of 0.89.
The metastable state is preferred given that $^{88}$Zr is 0+ and
capture of a thermal neutron should be s-wave and hence lead to a dominant production of the $1/2-$ (metastable) isomer. 
 
\section{Experimental Setup}
A sample of $^{88}$Zr was produced at Los Alamos National Laboratory Isotope Production Facility through proton irradiation of a yttrium target $^{89}$Y(p,2n)$^{88}$Zr and was delivered to the Nuclear Engineering Teaching Laboratory (NETL) at the University of Texas at Austin in 2N HCl solution. $^{88}$Zr electron capture decays to $^{88}$Y with a 83.4 day half-life which further decays through electron capture to stable $^{88}$Sr with a 106.6 day half-life. As the sample was over a year past initial separation, the $^{88}$Y activity exceeded the $^{88}$Zr activity. Furthermore, $^{88}$Y gammas, predominately 898 keV (93\% branching ratio) and 1.836 MeV (99\% branching ratio), provide a more significant background than the main 393 keV (97\% branching ratio) $^{88}$Zr gamma. Together, these factors motivated a $^{88}$Zr/$^{88}$Y separation. This was accomplished by immersing the sample within 0.01 M HDEHP in dodecane. The zirconium separates into the organic (HDEHP) layer while the yttrium remains in the aqueous (HCl) layer. The two liquids were then separated with a separatory funnel. 
Finally, the $^{88}$Zr in the organic solvent was extracted with 1.1M oxalic acid, pulling it into the aqueous layer. The 1.1M oxalic acid solution was digested with 2 M nitric acid and 1 M hydrogen peroxide at 100 $^\circ$C. Once the oxalic acid was digested, the $^{88}$Zr residue was dissolved in 2M nitric acid, transferred dropwise onto Kapton tape, and evaporated to dryness by heating. Kapton is a polyimide (C$_{41}$H$_{22}$N$_4$O$_{11}$) film with a silicone adhesive backing. The dried sample was sealed in two HDPE vials for irradiation along with a natural molybdenum flux monitor wire. A $^{88}$Zr sample activity of 5$\mu$Ci was chosen to maximize the number of target atoms without creating an unmanageable dead-time during post-exposure counting.

The simultaneous measurement of $^{89\textrm{m}}$Zr and $^{89}$Zr requires a high neutron flux as well as prompt counting after exposure. The TRIGA Mark II nuclear research reactor at NETL is well suited for this requirement. In addition to a thermal neutron flux of $1\times10^{12}$ n/cm$^2$/s, it is equipped with a pneumatic transfer system (tPNT) which enables rapid implantation and removal of samples in the reactor.
This sample, accompanied by a 6mg natural molybdenum flux monitor, was irradiated for 10 minutes. Upon end of irradiation the sample was counted in a HPGe with an ORTEC DSPEC 50 MCA. Measurement began 180 seconds after end of irradiation, at which point the 588 keV $^{89\textrm{m}}$Zr activity was 500 counts/s. 

\section{Analytical Methods}
\begin{figure}
    \centering
    \includegraphics[width=0.9\linewidth]{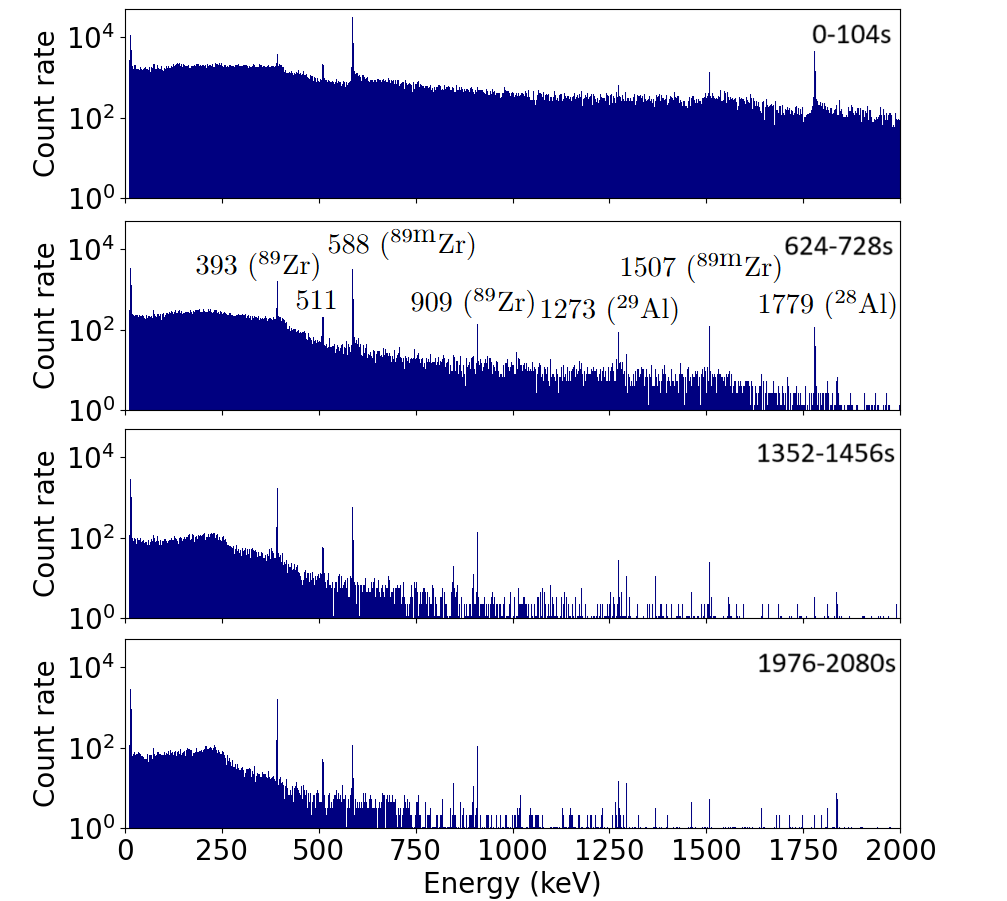}
    \caption{HPGe spectra recorded from sample, normalized for detector livetime and efficiency. Initial data taken 180s after end of irradiation.}
    \label{fig:init_spectrum}
\end{figure}

Consecutive spectra were extracted from the ORTEC MCA, corrected for deadtime, and background subtracted following the recipes of Gilmore \cite{Gilmore}. The expected gamma peaks were immediately apparent for $^{88}$Zr (393 keV), $^{89\textrm{m}}$Zr (588 keV and 1507 keV), and $^{89\textrm{g}}$Zr (909 keV). Other notable peaks included $^{28}$Al at 1780 keV with 2.2-minute half life.
Notably the 909 keV $^{89\textrm{g}}$Zr line was well-separated from the 898 keV $^{88}$Y line.
The time dependence of the 588 keV and 1507 keV $^{89\textrm{m}}$Zr signals is visible in Figure \ref{fig:init_spectrum}.

\begin{figure}
    \centering
    \includegraphics[width=0.9\linewidth]{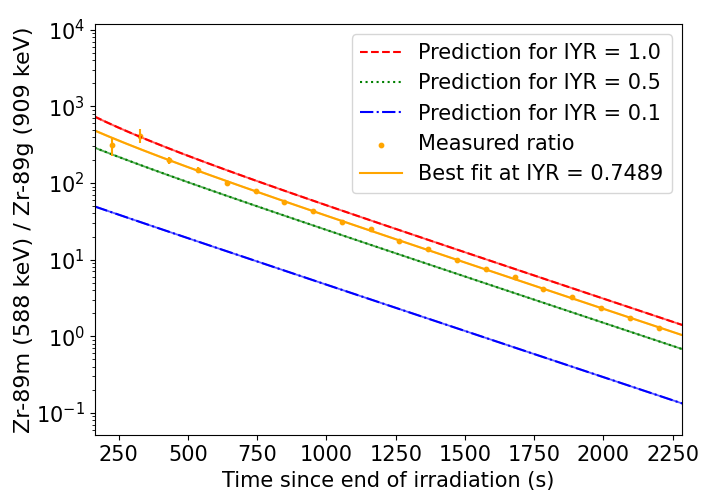}
    \caption{Best fit of IYR model to measured signal ratios. Predictions are shaded to indicate (negligible) uncertainty of calculations.}
    \label{fig:bestfit-plot}
\end{figure}

The amounts of $^{89}$Zr and $^{89\textrm{m}}$Zr created by the $^{88}$Zr(n,g) reaction both depend almost directly on flux, initial sample size, and cross section. By taking the ratio of the respective signals from these products, the uncertainty from these variables is minimized and the IYR better isolated. 
A numerical model was created which took into account $^{88}$Zr(EC)$^{88}$Y, $^{88}$Zr(n,$\gamma$)$^{89\textrm{g}/\textrm{m}}$Zr, $^{89\textrm{m}}$Zr(IT)$^{89\textrm{g}}$Zr, and $^{89\textrm{m}}$Zr($\epsilon/\beta^+$)$^{89}$Y, as well as the delay in the 909 keV $^{89\textrm{g}}$Zr line due to its path through the 16-second half-life $^{89\textrm{m}}$Y state.
Modelling of the evolution of the 588 keV to 909 keV ratio over time showed that the ratio changes with initial magnitude mostly dependent on the IYR and a slope dependent mostly on the half-life of $^{89\textrm{m}}$Zr.
Example predictions are shown in Figure \ref{fig:bestfit-plot}. Note that the flux and cross section cancel to first order, though they do not cancel to higher orders. For instance, the burnup of $^{88}$Zr depends on the flux and cross section and therefore varies the line intensity ratio in a slight manner. As such, terms like these are taken into account in the final fit.


A chi-squared fit was applied to determine the IYR value for best fit. Other free variables were added to the chi-squared function as penalty terms. The $\chi^2$ values are modified from the chi-squared of the fit ($\chi^2_\textrm{stat}$) according to:

\begin{equation}
\chi^2 = (\chi^2)_{\textrm{stat}} + \sum_{n=1} \frac{\Delta\alpha_i}{\sigma_{\alpha_i}}
\end{equation}

\noindent where $\Delta\alpha_i$ is the difference from the accepted value to to the best fit value and $\sigma_{\alpha_i}$ is the standard deviation of the accepted value, following the convention of the Particle Data Group \cite{PDG2020}. This allows the variables to be modified during the fitting process to a magnitude dependent on the precision of the variable's accepted value. This enabled the analysis to account for uncertainties in flux, cross section, and isotope half lives, as well as internal conversion rates which affect the gamma emission levels. The MINUIT package \cite{James1975}\cite{iminuit} was used to calculate the final fit by minimization of chi-square, with the migrad solver applied after the simplex minimizer.

The IYR value which provides a minimum chi-squared ($\chi^2_\textrm{min}$) is the measured value given here with 68\% confidence limits given by the range of IYR values that provide an increase in chi-squared of 1.0 ($\Delta\chi^2$).

\begin{equation}
\chi^2 = \chi^2_\textrm{min} + \Delta\chi^2
\end{equation}

\noindent In the same manner, 95\% confidence limits can be tabulated with $\Delta\chi^2=3.84$, etc \cite{PDG2020}.

The following inputs are allowed to float in the mathematical model. The $^{88}$Zr(n,$\gamma$)$^{89}$Zr cross section used is the 2021 Shusterman result of 804,000$\pm$63,000 barns \cite{Shusterman2021}. The flux is derived by scaling an MCNP spectra which has been tuned by 47 foils and subsequent unfolding with the SAND-II package \cite{Rising2023}\cite{Griffin1994} with the amount of $^{99}$Mo activity created during exposure of the natural molybdenum monitoring wire. The flux at 0.025 eV is 4.38$\pm$0.22$\times10^{12}$ $n/cm^2/s$. As the cross section and flux appear together in any relevant formula in the numerical model, they are treated as one systematic with a value of 3.52$\pm$0.33$\times10^{-6}$ $n/s$. The absolute emission rate of the 909 keV gamma from $^{89}$Zr decay is 0.9914$\pm$0.0012 \cite{Fenwick2020} due to internal conversion; the 588 keV gamma from $^{89\textrm{m}}$Zr appears at a rate of 0.9533$\pm$0.0007 for the same reason \cite{2008Ki07}. These coefficients, as well as the relative efficiency of the HPGe system (calibrated with $^{152}$Eu), are also combined whenever used in the numerical model; thus they are quoted together with systematic error in quadrature. Half lives of multiple isotopes are important parameters: $^{88}$Zr is 83.4$\pm$0.3 days \cite{Muller1988}, $^{89}$Zr is 78.361$\pm$0.025 hours \cite{Fenwick2020}, $^{89\textrm{m}}$Zr is 4.161$\pm$0.010 minutes \cite{NUBASE}, and $^{89\textrm{m}}$Y is 15.663$\pm$0.005 seconds \cite{NUBASE}. These are compiled in Table \ref{penalty-table}.

\section{Results}\label{sec:Results}

The half life of $^{89\textrm{m}}$Zr was measured at 247$\pm$5 s in agreement with the accepted value of 249.7$\pm$0.6 s \cite{NUBASE}. The relative rate of 1507 keV emissions (resulting from $^{89\textrm{m}}$Zr positron emission) to 588 keV emissions was measured at 14.59$\pm$0.22, in agreement with the previous result of 14.81$\pm$0.14 from Van Patter et al. \cite{VanPatter1964}.

The final fit achieved by MINUIT yields a value for the isomeric yield ratio of 0.7489$\pm$0.0061, with 95\% confidence range of $\pm$ 0.0129. 

Chi-squared minimization as performed by MINUIT resulted in deviations in other variables when identifying the best fit IYR value. These deviations are recorded in Table \ref{penalty-table}. The most significant is that for $^{89\textrm{m}}$Zr half life at 0.3 standard deviations. 

\begin{table}[h!]
		
		\begin{center}
\resizebox{6.0in}{!}{
\begin{tabular}{|c|c|c|}\hline 
Parameter&Value&Deviation at best-fit ($\sigma$) \\\hline
Cross section $\times$ flux&3.52$\pm$0.33$\times10^{-6}$ \textrm{n/s} \cite{Shusterman2021}& $<10^{-3}$ \\\hline
$^{88}$\textrm{Zr half-life} &83.4$\pm$0.3 \textrm{ d} \cite{Muller1988}& $<10^{-3}$ \\\hline
$^{89}$\textrm{Zr half-life} &78.361$\pm$0.025 \textrm{ h} \cite{Fenwick2020}& $< 10^{-3}$ \\\hline
$^{89m}$\textrm{Zr half-life} &4.161$\pm$0.010 \textrm{ m} \cite{NUBASE}& 0.33 \\\hline
$^{89m}$\textrm{Y half-life} &15.663$\pm$0.005 \textrm{ s} \cite{NUBASE}& $< 10^{-3}$ \\\hline
 \textrm{Rel. Eff. $\times$ ICC} &1.44671$\pm$0.00126 \cite{Fenwick2020}&  0.002 \\\hline
$^{89m}$\textrm{Zr branching ratio} &0.9377$\pm$0.0012 \cite{NUBASE}& $< 10^{-3}$ \\\hline

\end{tabular}
}

\end{center}
\caption{Fit parameter values and corresponding penalty term deviations at best MINUIT fit.}
\label{penalty-table}
\end{table}

\section{Conclusions}

This experiment has succeeded in measuring a fundamental aspect of the $^{88}$Zr neutron capture reaction. The IYR is 0.7489$\pm$0.0061.
We note that TALYS with $^{89}$Zr energy levels given by the RIPL-3 database gives an IYR of 0.89.
Furthermore, the IYR was assumed to be unity in the seminal paper discovering the large $^{88}$Zr(n,$\gamma$) cross section \cite{Shusterman2019}. In this paper, the cross section was determined both by measuring $^{88}$Zr depletion and $^{89}$Zr conversion. We note that an additional 1.5\% of $^{89}$Zr conversion should have been predicted to occur since 25\% of $^{88}$Zr(n,$\gamma$)$^{89}$Zr proceeds directly to the ground state and avoids the 6.2\% probability of decay to $^{89}$Y rather than isomeric transition. The authors also note that $^{89\textrm{m}}$Zr half-life and branching ratios measured here agree with previous measurements.

\section{Acknowledgements}

We thank our collaborators and group members, especially Genevieve Alpar, Said Bakkar, David Catlett, Emilio Mendoza Cembranos, Andrew Kelly, Albert Parmenter, Paul Phillips, Tracy Tipping, and Benjamin Vines for their valuable assistance throughout this experiment. The authors would also like to thank Jennifer Shusterman and Nicholas Scielzo for their encouragement and useful discussions throughout this work as well as Tom Carroll for his helpful insights on penalty term fitting. This work was supported by the University of Dallas Donald A. Cowan Physics Institute.



\begin{thebibliography}{1}

\bibitem{Shusterman2019}
J.~Shusterman \textit{et al.}
``The surprisingly large neutron capture cross-section of $^{88}$Zr''
Nature \textbf{565}, 328-330 (2019).
https://doi.org/10.1038/s41586-018-0838-z

\bibitem{Shusterman2021}
J.~Shusterman \textit{et al.}
``Aqueous harvesting of $^{88}$Zr at a radioactive-ion-beam facility for cross-section measurements''
Phys. Rev. C \textbf{103}, 024614 (2021).
https://doi.org/10.1103/PhysRevC.103.024614

\bibitem{Stamatopoulos2023}
A.~Stamatopoulos \textit{et al.}
``Discovery of the origin of the enormous 88Zr neutron-capture cross section and quantifying its impact on applications''
Phys. Rev. L Accepted (2023).
https://doi.org/10.21203/rs.3.rs-3331910/v1

\bibitem{Weiss2015}
C.~Weiss \textit{et al.}
``The new vertical neutron beam line at the CERN n$\_$TOF facility design and outlook on the performance''
NIM A \textbf{799}, 90-98 (2015).
https://doi.org/10.1016/j.nima.2015.07.027.

\bibitem{VanPatter1964}
D.~Van~Patter and S.~Shafroth
``Decay of 78.4 h Zr89 and 4.18 min Zr89m''
Nuclear Physics \textbf{50} 113-135 (1964).
https://doi.org/10.1016/0029-5582(64)90196-8

\bibitem{Mendoza2020}
E.~Mendoza \textit{et al.}
``NuDEX: A new nuclear $\gamma$-ray cascades generator''
EPJ Web of Conferences \textbf{239} 17006 (2020)
https://doi.org/10.1051/epjconf/202023917006

\bibitem{Mangal1963}
S.K.~Mangal and P.S.~Gill
``Isomeric cross section ratios for (n,2n) reactions at 14 MeV''
Nuclear Physics \textbf{49} 510-514 (1963).
https://doi.org/10.1016/0029-5582(63)90115-9

\bibitem{Katz1953}
L.~Katz, R.G.~Baker, R.~Montalbetti
``The photoneutron cross sections of Rb$^{87}$, Zr$^{90}$, and Mo$^{92}$''
Canadian Journal of Physics \textbf{31} 250-261 (1953).
https://doi.org/10.1139/p53-026

\bibitem{Satheesh2011}
B.~Satheesh, M.M.~Musthafa, B.P.~Singh, and R.~Prasad
``Nuclear isomers $^{90m,g}$Zr, $^{89m,g}$Zr, $^{89m,g}$Y, and $^{85m,g}$Sr formed by bombardment of $^{89}$Y with protons of energies from 4 to 40 MeV''
International Journal of Modern Physics E \textbf{20} 2119-2131 (2011).
https://doi.org/10.1142/S0218301311019702

\bibitem{TALYS}
A.J.~Koning, S.~Hilaire, S.~Goriely
``TALYS: modeling of nuclear reactions''
Eur. Phys. J. A \textbf{59} 131 (2023).
https://doi.org/10.1140/epja/s10050-023-01034-3


\bibitem{RIPL3}
R.~Capote \textit{et al.}
``Reference Input Parameter Library (RIPL-3)''
Nuclear Data Sheets \textbf{110} \textbf{12} 3107-3214 (2009)

\bibitem{Rising2023}
M.E.~Rising \textit{et al.}
``MCNP Code Version 6.3.0 Release Notes''
Los Alamos National Laboratory Tech. Rep. LA-UR-22-33103 (2023).
https://doi.org/10.2172/1909545

\bibitem{Griffin1994}
P.J.~Griffin, J.G.~Kelly, J.W.~VanDenburg
``User's Manual for SNL-SAND-II Code''
Los Alamos National Laboratory Tech. Rep. SAND93-3957 (1994).
https://doi.org/10.2172/10149711


\bibitem{Muller1988}
H.-W.~Müller
Nuclear data sheets for A = 88,
Nuclear Data Sheets,
Volume 54, Issue 1,
1988,
Pages 1-97,
ISSN 0090-3752,
https://doi.org/10.1016/S0090-3752(88)80107-8.

\bibitem{Fenwick2020}
Andrew~J.~Fenwick \textit{et al.}
``Absolute standardisation and determination of the half-life and gamma emission intensities of 89Zr''
Applied Radiation and Isotopes \textbf{166} 109294 (2020)
https://doi.org/10.1016/j.apradiso.2020.109294

\bibitem{KaZm1992}
K.~Kawade \textit{et al.}
Measurement of Beta-Decay Half-Lives of Short-Lived Nuclei,
JAERI-M,
92-027,
Pages 364-368

\bibitem{2008Ki07}
T.~Kibédi, T.W.~Burrows, M.B.~Trzhaskovskaya,P.M.~Davidson, C.W.~Nestor,~Jr. 
'Evaluation of theoretical conversion coefficients using BrIcc' Nucl. 
Instr. and Meth. A 589 (2008) 202-229

\bibitem{NUBASE}
G.~Audi, O.~Bersillon, J.~Blachot, A.H.~Wapstra, ``The Nubase evaluation of nuclear and decay properties,''
Nuclear Physics A,
Volume 729, Issue 1,
2003,
Pages 3-128,
ISSN 0375-9474,
https://doi.org/10.1016/j.nuclphysa.2003.11.001.

\bibitem{James1975}
F.~James, M.~Roos
``Minuit - a system for function minimization and analysis of the parameter errors and correlations''
Computer Physics Communications \textbf{10} 343-367 (1975).
https://doi.org/10.1016/0010-4655(75)90039-9

\bibitem{iminuit}
H.~Dembinski \textit{et al.}
``scikit-hep/iminuit Python Interface''
https://doi.org/10.5281/zenodo.3949207

\bibitem{Gilmore}
G.~Gilmore
``Practical Gamma-ray Spectrometry'' 2nd edition.
Wiley, Chichester
2008

\bibitem{PDG2020}
P.A.~Zyla \textit{et al.} (Particle Data Group)
``The Review of Particle Physics''
Prog. Theor. Exp. Phys. \textbf{2020} 083C01 Chapter 40 (2020).
https://doi.org/10.1093/ptep/ptaa104

\end{thebibliography}
\end{document}